# GPRS Based Intranet Remote Administration GIRA

[1]Shashi Kumar N.R. [2] R Selvarani [3]Pushpavathi T.P.

[1,3]MTech final year, CSE Dept. DSCE
chandravalli@yahoo.co.in, acepushpa@yahoo.co.in
[2]Head Software Engineering Research - RIIC, CSE Department, selvss@yahoo.com
[1,2,3] Dayananda Sagar Institutions, Bangalore, India

**Abstract**

In a world of increasing mobility, there is a growing need for people to communicate with each other and have timely access to information regardless of the location of the individuals or the information. With the advent of moblle technology, the way of communication has changed. The gira system is basically a mobile phone technology service. In this paper we discuss about a novel local area network control system called gprs based Intranet Remote Administration gira. This system finds application in a mobile handset. With this system, a network administrator will have an effective remote control over the network. gira system is developed using gprs, gcf Generic Connection Framework of j2me, sockets and rmi technologies.

## System Hypothesis

The GIRA system enables a Network Administrator to monitor and control the intranet through a mobile device. Remote Administration of intranet facilitates in getting the list of clients on network, control the status of the PC (abort or run), and to execute certain applications, chat with clients over network and broadcast message etc.

## I. Introduction

Consumers of mobile and small computing devices have high performance expectations. They demand quick response time, compatibility with companion services and full-featured applications in a small device. Consumers expect the same software and capabilities found on their desktop and laptop computers to be available on their cell phones and personal digital assistants [1]. To meet these expectations, developers have to rethink the way they build a mobile device. Developers need to harness the power of existing front-end and back-end software found on business computers and transfer this power onto small, mobile and wireless computing devices. J2ME enables this transformation to occur with minimal modifications, assuming that applications are scalable in design so that an application can be custom-fitted to resources available on a small computing device [1]. Developers must determine the minimum client processing that will meet the end user's expectations of quick response time that is feasible within the limited resources available on the small computing device [1]. Traditional computing devices use fairly standard hardware configurations such as a display, keyboard, mouse and large amounts of memory and permanent storage. However the new breed of computing devices lack hardware configuration continuity among devices. Some devices do not have a display, permanent storage, keyboard or mouse. And memory availability is inconsistent among small computing devices [1].

The existing system allows Systems Administrator to monitor and control the Intranet or Local Area Network only from the place where it is installed or from where the network node extends and exists. The proposed system allows a Network Administrator to monitor and control the Intranet or Local Area Network from the existing location as well as from a remote location using mobile device enabled with GPRS.

## II. Motivation

Wireless communication is one of the biggest research areas among recent emerging technologies. Handheld devices such as a mobile phone require certain applications and services for efficient utilization of the device. Since mobile phones, as the name indicates are mobile in nature, services related to monitoring and remote controlling of Local Area Network will be more useful. The above issues motivated the study and implementation of GPRS based monitoring and controlling of Local Area Network.

## III. Mobile Computing

The changing wireless market is, driven by customers who want access to services and applications that will add value to their leisure and work, and by operators who need a return on their huge investments in 3G licenses and infrastructure. For an ordinary person conducting business on the go, there are often many tasks that need to be completed in a given day. Some of these tasks must be completed at a particular time, at a particular location. For instance, a Network Administrator may need to monitor and control the local area network from home or from the car he is driving i.e., when he is not in the office. He wants to monitor and control the network when out of office. How is this possible? It is possible by selecting the right technology for developing a system to Monitor and Control the Intranet. After the selection of technologies we need to integrate them to achieve the desired results.

## IV. Feasibility Study

**Technical Feasibility** Technical feasibility is the hardware and software requirements which are needed to implement the proposed system in the organization. Technical requirements need to be fulfilled to make the proposed system work. This should be predetermined to make the system more competent. The system is technically feasible because it is implemented with J2ME, Servlets, RMI and Socket programming concepts which are the proven technologies.

**Economical Feasibility** The Economical feasibility must satisfy the needs of the technical feasibility and the operational feasibility. It involves the economic feasibility of developing and implementing the proposed system. The GIRA system is economically feasible as the system comprises only the GPRS enabled Mobile phone and an existing LAN setup. Hence the system is less expensive.

**Operational Feasibility** The designed GIRA system is easy to use for any normal user to perform any task with minimal knowledge of the system.

## V. System Design Implementation

The system has been implemented by developing an application for a mobile device using J2ME,designing the interface between a mobile device and server using GCF, GPRS, and Tomcat Web server, establishing connection between a mobile device and clients on the intranet using RMI and Sockets server. This plays a centralised role by controlling all the activities on the intranet. The clients are connected to mobile through the RMI/Sockets.

**Functional Description**

This system facilitates the administrator to provide maximum details about the network through a mobile device remotely. After giving the correct login name, password along with the proper IP address of the server in the login form, the administrator can make a choice from the menu. If the login information given is not proper then the user will not be able to connect to the server. Some of the features are as follows: Connection between a mobile user and LAN is established using GCF, GPRS and a Tomcat Web server. GCF is a Generic Connection Framework, a J2ME solution to establish network connection. Generic Connection Framework creates connection between small computing device and a remote computer. GCF uses http, ftp, and socket communication protocols.

***Connection establishment***

```
String
stral="http://"+ServerIPAddress.getServerIPAddress()
+":80/process/servlet/"+cname+"?"+message;
con=(StreamConnection)Connector.open(stral,Connec
tor.READ_WRITE);
InputStream in=con.openInputStream();
```

After getting the connection to a server IP address, the mobile user has to send his authentication details. This will be checked and verified by the database of the server using JDBC as follows:

```
try
   {
     Class.forName("sun.jdbc.odbc.JdbcOdbcDriver");
          Connection
con=DriverManager.getConnection("jdbc:odbc:users")
;
          Statement stmt=con.createStatement();
          StringTokenizer st=new
StringTokenizer(msgCommand,"&");
          st.nextToken();
          ResultSet   rs=stmt.executeQuery("select   *
from users where username='"+st.nextToken()+"' and
password='"+st.nextToken().trim()+"'");
          if(rs.next())
          {
message="Logged In";
           login=true;
          }
          else
```

```
            {
              message="Invalid user";
            }
            con.close();
                    }
```

Clients Waiting Process and service request:

```
System.loadLibrary("process");
java.rmi.registry.LocateRegistry.createRegistry(1099);
Naming.bind("client",new client());
    try
{
      ServerSocket ss=new ServerSocket(7070);
      while(true)
          {
        Socket so=ss.accept();
        new ChatWindow(so);
      }
    }
    catch(Exception e){e.printStackTrace();}
  }
```

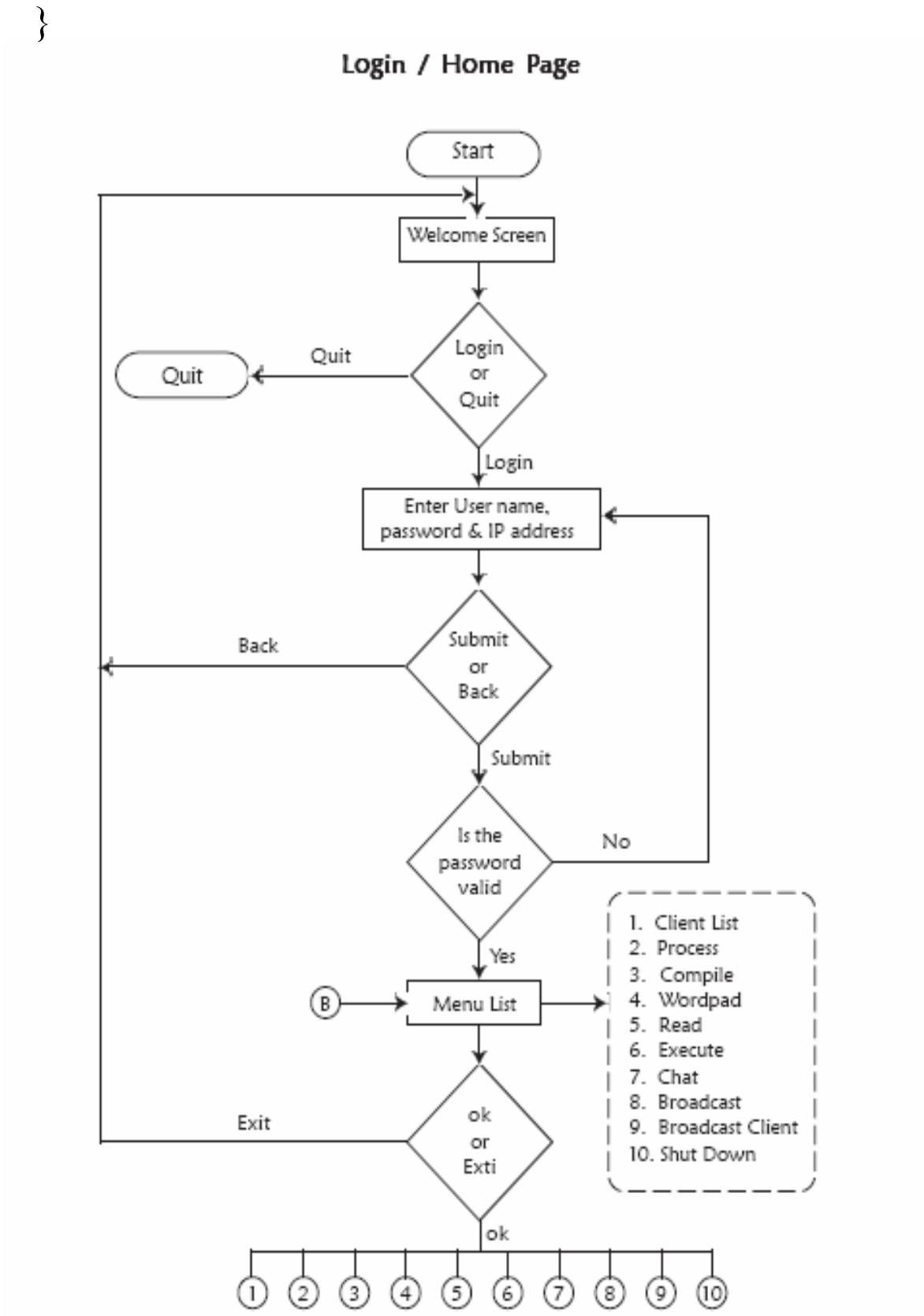

**Figure 1** Flowchart-Main menu and Options

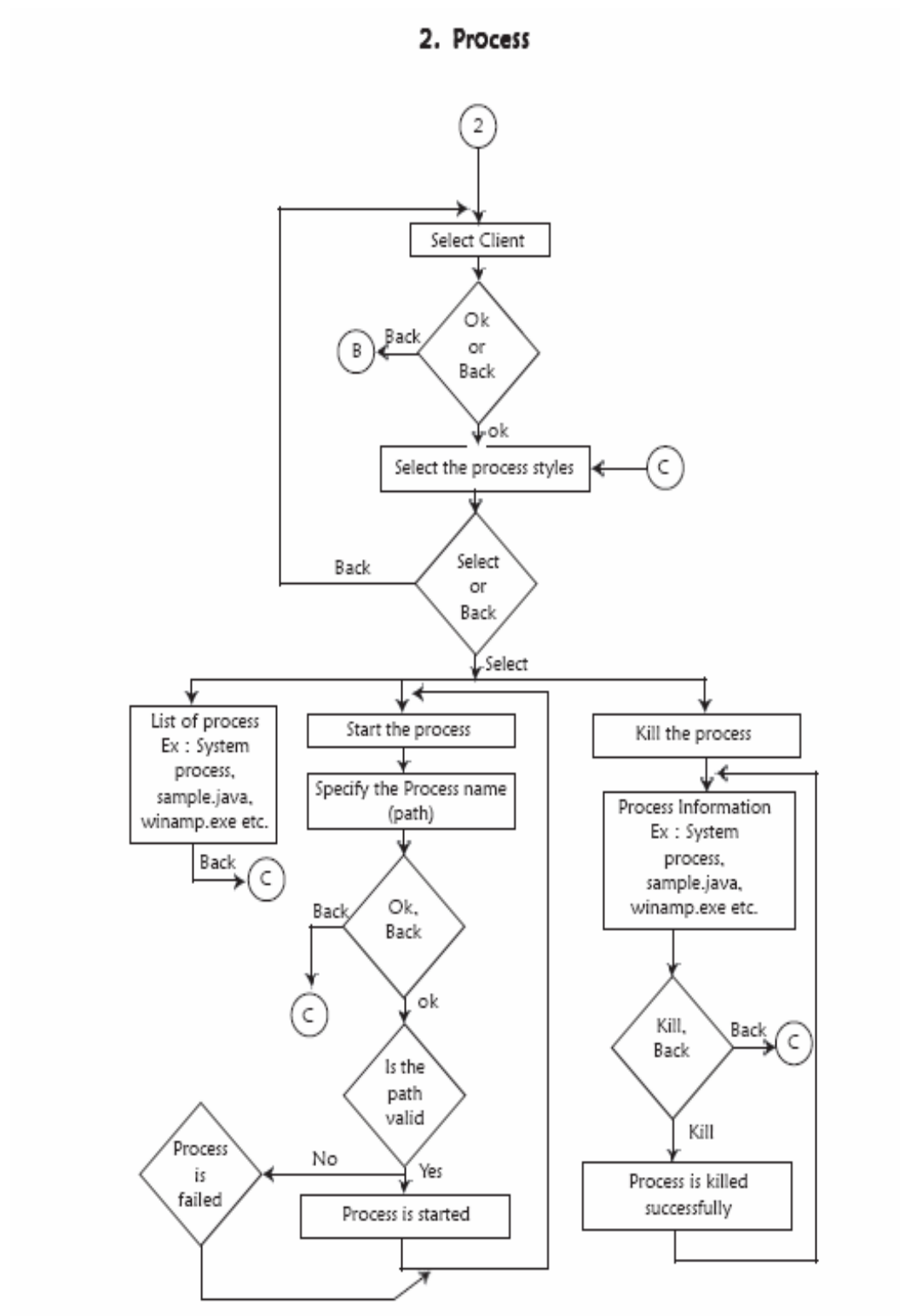

**Figure 2** flow chart for Option 2: Control Processes

1. **Client List** An administrator will be able to view the names of all the clients connected to the server along with the clients.
2. **Process** An administrator will be able to choose any one of the clients among the given list, and he can choose the options like 1) List of process 2) Start the process 3) Kill (abort) the process.
3. **Compile** An administrator can compile a file he wishes to do from a mobile device after specifying the path and file name.
4. **WordPad** A user can type a text and save into a file of a particular client.
5. **Read** A user will be provided with all the drives existing in a specific client and any file of a particular drive can be opened for read in a mobile device.
6. **Execute** This menu facilitates a user from compiling any java programs from a mobile device after explicitly mentioning a particular java file that exists in a particular drive of a specific client.

7. **Chat** This option enables the administrator to have on line interaction with any of his clients.
8. **Broadcast** This option facilitates an administrator for broadcasting a message to all the clients and a server from a mobile device.
9. **Client Messaging** This option provides the facility to the administrator to send messages to a particular client.
10. **Shut Down** This option closes the entire operation.

## VI. Conclusion and Future Work

In this Paper we have presented a novel LAN remote administration system called GIRA, which is based on the technological advancement of GPRS, RMI and Socket programming. We have focused on implementation of connectivity between a mobile device and an intranet using GPRS and Generic Connection Framework (GCF) of J2ME which provides an extensible, generic I/O framework for resource constrained devices. The work also includes RMI implementation as it allows object-to-object communication and their execution for resource sharing between different systems. The GIRA system also uses Socket programming as it allows lower level of abstraction for network programming to make two way communications between a mobile user and any client system on a LAN. Dynamic Link Libraries are created for using OS level operations like remote shutdown of the system. This System is designed for windows operating system for system level operation like shutdown. In future our work may be extended to system level operation for Linux operating system.